\documentclass[aps,prl,twocolumn,longbibliography,superscriptaddress,amsmath,amssymb,floatfix,showpacs]{revtex4-2}
\usepackage{amsmath}
\usepackage{amssymb}
\usepackage{bm}
\usepackage{color}
\usepackage[utf8]{inputenc}
\usepackage{pifont}
\usepackage[colorlinks,linkcolor={blue},citecolor={blue},urlcolor={blue}]{hyperref}
\usepackage{mathtools}
\usepackage{booktabs}
\usepackage{multirow}
\usepackage{physics}
\usepackage{relsize}
\usepackage{mathrsfs}
\usepackage{mathrsfs}
\usepackage{mathdots}
\usepackage{babel}
\begin{document}

\title{Photo-induced Multiply Quantized Vortex States in Dirac-like Materials}

\author{Lauren I. Massaro}
\affiliation{Department of Physics, Kennesaw State University, Marietta, Georgia 30060, USA}
\author{Connor Meese}
\affiliation{Department of Physics, Kennesaw State University, Marietta, Georgia 30060, USA}
\author{Nancy P. Sandler}
\affiliation{Department of Physics and Astronomy and Nanoscale and Quantum Phenomena Institute, Ohio University, Athens, Ohio 45701, USA}
\author{Mahmoud M. Asmar}
\email{masmar@kennesaw.edu}
\affiliation{Department of Physics, Kennesaw State University, Marietta, Georgia 30060, USA}

\begin{abstract}
Subjecting a massive two-dimensional Dirac material to a vortex light beam provides a mechanism for the photo-induction of multiply quantized vortices. Using Floquet theory, we show that electronic vortices, characterized by their total angular momentum, are exclusive to circularly polarized vortex beams. The equations for the driven system at the one photon-resonance are mapped to the Bogoliubov–de Gennes equations of $s$-wave superconductors with multiply quantized vortices. This mapping provides valuable analytical tools for the analysis of the system's spectral properties.  
\end{abstract}
\maketitle

Dynamical control of materials' properties and phases through time-periodic modulations is a powerful tool for accessing, controlling, and generating quantum states of matter. This approach, known as Floquet engineering~\cite{flreview1,flreview2}, has become an active field of research as it provides non-thermal pathways to properties' control through light-matter interactions~\cite{flreview3,flreview6,flreview5}. Irradiation of solids gives rise to non-perturbative photon-dressed bands and quasiparticle excitations. The formation of Floquet--photon-dressed--bands has been observed on the surface of topological insulators via time-resolved and angle-resolved photoemission spectroscopy~\cite{FloqExp2,FloqExp1,FloqExp4}. Additionally, the formation of light-induced topological gaps has been measured through the anomalous Hall conductance in graphene~\cite{FloqExp3}. 

Optical tuning of phases of matter has been conventionally realized by light’s intensity, frequency, and polarization variations~\cite{flreview1,flreview2,flreview3,flreview6,flreview5}. However, these factors do not explore the full versatility light has to offer for the control of quantum matter. Additional degrees of tunability can arise from the spatial control of optical beams. Vortex light beams (VLBs) are examples of such sources of radiation~\cite{OVB-Allen,OVB-Review}. These beams carry an orbital angular momentum (OAM) in addition to the intrinsic spin angular momentum resulting from their circular polarization~\cite{lightV3,OVL-Oneil}. Recently, experiments investigating VLB-matter coupling in solids have detected OAM induced electric signals in GaAs~\cite{lightVexp9}, OAM dependent photocurrents in WTe$_2$ photodetectors~\cite{lightVexp10}, and an OAM enhanced photovoltaic effect in MoS$_2$~\cite{lightVexp11}.

\begin{figure}[ht!]
  \centering
  \includegraphics[width=0.5\textwidth]{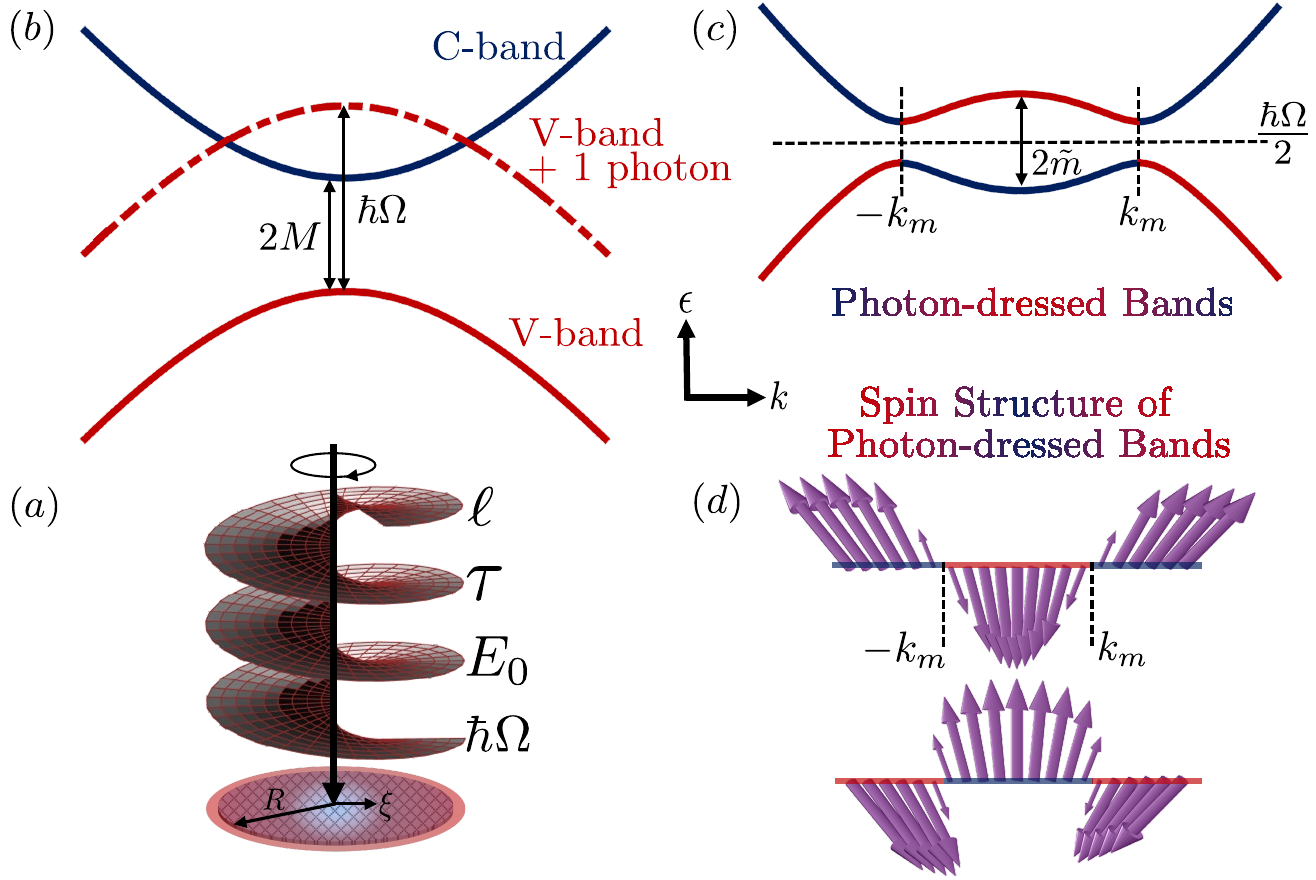}
  \caption{(a) Schematic representation of CP ($\tau$) monochromatic ($\hbar\Omega$) VLB carrying OAM ($\ell$) and amplitude $E_0$ normally incident to a massive Dirac material characterized by a length scale $R$. (b)-(c) Two-band Floquet model. The hole band (V-band) is dressed with one photon and intersects the electron band (C-band) at $\hbar v_{\rm F}|k_{m}|=\sqrt{(\hbar\Omega/2)^2-M^2}$. The hybridization at $\pm k_m$ opens dynamical gaps. (d) Spin configurations of Floquet states in the photon-dressed bands in (c).}
  \label{Fig1}
\end{figure}

In this paper, we demonstrate the photo-induction of multiply quantized vortices in two-dimensional massive Dirac materials. Utilizing Floquet theory, we show that a VLB-driven massive Dirac material hosts multiply quantized vortex states characterized by their total angular momentum {\it if and only if} the VLB is circularly polarized. Moreover, we show that the Floquet-Hamiltonian equations governing the driven system can be mapped, in the one-photon resonance regime, to Bogoliubov–de Gennes equations for $s$-wave superconductors or fermionic superfluids under applied magnetic flux manifesting multiply quantized vortices~\cite{multiply}, which allows insights into the connection between these seemingly unrelated systems.  

The mechanism for vortex generation in matter via light-driving can be illustrated in a simple form by considering a material hosting massive Dirac-like quasiparticles subjected to a monochromatic VLB
which is described by the spatio-temporal Hamiltonian
\begin{eqnarray}\label{t-dep-H} 
H({\bm r}, t)= v_{\rm F}{\bm \sigma}\cdot {\bm p}+M\sigma_z+ e v_{\rm F} {\bm{\mathcal{{A}}}}({\bm r, t}) \cdot {\bm\sigma}\;. 
\end{eqnarray}
Here, $v_{\rm F}$ is the quasielectron's Fermi velocity, $2M$ is the electronic bandgap, $e>0$ is the electron charge, and ${\bm \sigma}$ is a vector of Pauli matrices acting on the electronic spin. The vector potential describing the VLB is ${\bm{\mathcal{{A}}}}({\bm r, t})=A(r)\Re[e^{i(\Omega t -\ell\theta)}\varepsilon_{\eta}]$, where $\theta$ is the azimuthal angle and $\Omega$ is the light's frequency. $\varepsilon_{\eta}=\hat{x}+e^{-i\eta}\hat{y}$ and $\eta \in [0,2\pi)$ determine the light's polarization, {\it e.g.}, by varying $\eta$ from $0$ to $\pi/2$ we tune the polarization from linear to circular. VLBs are characterized by a spatial dependence $A(r)$ that vanishes at $r=0$ for non-zero values of OAM $\ell$, ensuring its vortex structure~\cite{OVB-Allen,OVB-Review,lightV3,OVB-Review}, as seen in Fig.~\ref{Fig1} (a). We also consider a paraxial beam~\cite{OVB-Review} with a characteristic length $\xi$ which is smaller than the length scale $R$ defining the system, Fig.~\ref{Fig1} (a). The amplitude of the beam varies smoothly and saturates for radii beyond $\xi$ to $A_0=E_0/\Omega$ where $E_0$ is the amplitude of the light's electric field component. We also assume that the amplitude of the envelope function of typical Laguerre-Gaussian VLBs~\cite{OVB-Review} falls beyond the sample edges. Hence, a convenient parametrization of the $r$-dependence of the beam is $A(r)=A_0\tanh(r/\xi)$.

The time periodicity of $H({\bm r},t)$ in Eq.~\eqref{t-dep-H} grants the time-dependent Schr\"{o}dinger Equation, $[H({\bm r}, t)-i\hbar\partial_t]\hat{\psi}({\bm r},t)=0$, solutions of the form $\hat{\psi}_n({\bm r},t)=e^{-i\epsilon_nt/\hbar}\hat{\phi}_n({\bm r},t)$ through Floquet theory~\cite{Floq-Shirley}, where the spinor $\hat{\psi}^{{\rm T}}_n({\bm r},t)=[\psi_{n,\uparrow}({\bm r},t), \psi_{n,\downarrow}({\bm r},t)]$, $n\in\mathbb{Z}$ labels the Floquet modes and $\epsilon_n$ is the quasienergy modulo $\hbar\Omega$. The periodicity of the Floquet states $\hat{\phi}_n(\bm{r},t+2\pi/\Omega)=\hat{\phi}_n(\bm{r},t)$ facilitates their Fourier series representation $\hat{\phi}_n(\bm{r},t)=\sum_{m=-\infty}^{\infty}e^{i m \Omega t} \hat{\phi}^{m}_{n}({\bm r})$, allowing the expansion of the conventional Hilbert space to the Floquet space~\cite{Floq-Sambe}. In the Floquet space basis, $\hat{\phi}^{\rm T}_n({\bm r})=[\dots, \hat{\phi}^{1}_n({\bm r}),\hat{\phi}^{0}_n({\bm r}),\hat{\phi}^{-1}_n({\bm r}),\dots ]$, the Floquet Hamiltonian $H^{\rm F}=H({\bm r}, t)-i\hbar\partial_t$ is time-independent and takes the block-tridiagonal form 
\begin{equation}\label{H_Floq}
H^{{\rm F}}_{m'm}({\bm r})= [H_{0}(\bm r) + m\hbar\Omega]\delta_{m',m} + \sum_{\varsigma=\pm}H_{\varsigma}(\bm r)\delta_{m',m+\varsigma}\;.
\end{equation} 
$H_{0}(\bm r)$ is the $t$-independent part of Eq.~\eqref{t-dep-H}, and we define, $H_{\varsigma}(\bm r)={\bm v}^{\dag}_{\varsigma}(\bm r)\cdot {\bm \sigma}_{\varsigma}$, ${\bm v}^{\dag}_{\varsigma}=[\tilde{A}(r)e^{-i\varsigma \ell \theta}/2](1-i\varsigma e^{-i\varsigma\eta},1+i\varsigma e^{-i\varsigma\eta})$, $\tilde{A}(r)=ev_{\rm F}A(r)$, ${\bm \sigma}^{\dag}_{\varsigma}=(\sigma_{-\varsigma},\sigma_\varsigma)$, and $2\sigma_{\varsigma}=\sigma_x+i\varsigma\sigma_y$.  

Under monochromatic driving the electronic bands are photon-dressed and form Floquet bands. The Floquet quasienergy spectrum can thus be understood by overlaying replicas of the equilibrium Hamiltonian that are shifted by integer multiples of $\hbar\Omega$. The relative strength of the driving frequency to the gap-size of the material defines three main driving regimes: Off-resonance, below-resonance, and resonant (on/above resonance)\cite{flreview1,flreview2}. 

A massive Dirac material can be driven off-resonance for $\hbar\Omega\gg W$ where $W$ is the material's bandwidth, and hence, states in the conduction and valence bands do not couple. In this case, we can find an effective $2\times2$ Hamiltonian, using the van Vleck high-frequency approximation~\cite{VanVleck}, that describes the stroboscopic evolution of the system at each period~\cite{Virtual-ph,graphene-top-ins}. Up to first order in $(\hbar\Omega)^{-1}$ this effective Hamiltonian is $H^{\rm v}_{\rm eff}(\bm r)=H_{0}(\bm r)-\tilde{A}^2(r)\sin(\eta)\sigma_{z}/(\hbar \Omega)$. Hence, the effect of off-resonant VLB irradiation is the induction of a mass that depends on the distance from the vortex center, reaching maximum and minimum values as the polarization varies from linear to circular. The light-induced mass term originates from virtual photon processes in which every photon is absorbed and emitted~\cite{Virtual-ph,flreview2}, nullifying the OAM transferred to the quasielectrons. Therefore, the light-induced perturbation in this regime is OAM independent and does not create in-gap vortex states.     

In the resonant regime, however, overlapping Floquet replicas hybridize and gaps open at the location of bands' crossings in the Floquet spectrum~\cite{OnePh1}. Consequently, unlike the off-resonant regime where no new spectral gaps are generated, the resonant driving regime, defined by $2M<\hbar\Omega< W$, allows for the formation of light-induced gaps. For instance, as seen in Figs.~\ref{Fig1} (b) and (c), when a massive Dirac material is subjected to a space-homogenous CP light, the one-photon-dressed valence band intersects the conduction band at momenta $\pm k_{m}$ and a gap opens at the one-photon resonance, $\hbar\Omega/2$ (edge of Floquet Brillouin zone). Due to the strong band renormalization which results from electron-hole hybridization through the absorption of one photon, we can anticipate that the OAM carried by the VLB will have striking consequences on the light-induced states of the system.   

To capture the VLB induced spectral proprieties and characterize the states near the one-photon resonance $\hbar\Omega/2$, we consider an approximate two-band Floquet Hamiltonian~\cite{graphene-top-ins,OnePh1,OnePh2,OnePh3} that consists of the one and zero photon sectors of the full Floquet Hamiltonian in Eq.~\eqref{H_Floq}:
\begin{equation}\label{effectiveH}
  H^{\rm F}_{\rm eff}(\bm r)=\left[H_0(\bm r)+\frac{\hbar\Omega\sigma_0}{2}\right]\alpha_{0}+\frac{\hbar\Omega\sigma_0}{2}\alpha_z+\sum_{\varsigma}H_{\varsigma}(\bm r)\alpha_{\varsigma}\;.
\end{equation}
Here, the Pauli matrices $\alpha$ act on the $m'=\{1,0\}$ Floquet subspace. This Hamiltonian accurately captures the system's properties for a light-matter coupling $g=ev_{\rm F}A_{0}/(\hbar\Omega)\ll1$ where additional anti-crossings do not impact the quasienergy spectrum~\cite{graphene-top-ins,OnePh1,OnePh2,OnePh3,supp}. 
  
The angular and radial dependence of the effective Floquet Hamiltonian suggests the existence of a conserved generalized total angular momentum operator. To find its expression, we recall that at equilibrium, the angular momentum defined by $J^{{\rm eq}}_{z}=\mathcal{L}_z\sigma_0+\sigma_z/2$ (in units of $\hbar$) is a symmetry of $H_0(\bm r)$ [Eq.~\eqref{H_Floq}]. Here, $\mathcal{L}_z=-i\partial_\theta$, accounts for the $z$-component of the orbital angular momentum in real space, with eigenvalue $\ell_e$, and $\sigma_z/2$ for the  intrinsic spin angular momentum~\cite{AsmarTr2,AM1,AsmarTr4}. It is easy to verify that $J^{{\rm eq}}_{z}\alpha_0$ is not a symmetry of the projected Floquet Hamiltonian, as this operator does not account for the orbital and spin contributions introduced by the VLB. Consequently, one must extend the definition of the angular momentum operator $\mathcal{J}^{\rm F}_{z}$, to the Floquet space. To find it we consider the general ansatz $\mathcal{J}^{\rm F}_{z}=\mathcal{L}_z\sigma_0\alpha_0+h(\eta,\ell)\sigma_z\alpha_0/2+f(\eta,\ell)\sigma_0\alpha_z/2$, where $h(\eta,\ell)$ and $f(\eta,\ell)$ are arbitrary functions of the light's polarization $\eta$ and the OAM $\ell$. Requiring that $[\mathcal{J}^{\rm F}_{z},H^{\rm F}_{1,0}(\bm r)]=0$, we find that the function $h(\eta,\ell)=1$ while the function $f(\eta,\ell)$ must simultaneously satisfy $(f-\ell+\tau)(1\pm ie^{\pm i\tau\eta})=0$, where $\tau=\pm1$. These equations are satisfied concomitantly for $\eta=\tau\pi/2$ and $f=\ell+\tau$. As a consequence, there exists a total angular momentum operator that is conserved by the projected Floquet Hamiltonian {\it if and only if} the VLB is circularly polarized (CP). This operator is given by
\begin{equation}\label{Jz}
\mathcal{J}^{\rm F}_{z}=\mathcal{L}_z\sigma_0\alpha_0+\frac{\sigma_z\alpha_0}{2}+(\ell+\tau)\frac{\sigma_0\alpha_z}{2}\;,
\end{equation}
where $\tau$ determines the circular polarization handedness. Because, $\mathcal{J}^{\rm F}_{z}$ is a symmetry of the non-perturbative Floquet Hamiltonian in Eq.~\eqref{H_Floq} as we show in~\cite{supp}, the expression given in Eq.~\eqref{Jz} is the $m'=\{1,0\}$ projection of the generalized operator.
 
These results are in striking contrast with those reported in Refs.~\cite{MD1,MD2}, where it was found that linearly polarized VLBs can conserve total angular momentum. We attribute the discrepancy to the momentum approximations in the effective Hamiltonian used in these works, that give a fixed spin direction for all states across the bands, thus artificially enlarging the system's symmetries. However, as seen in Fig.~\ref{Fig1} (d) and Eq.~\eqref{t-dep-H}, the system with and without driving has a non-trivial spin momentum coupling which constraints its symmetries and limits the conservation of $\mathcal{J}^{\rm F}_{z}$ to only CP VLBs.    

Next, we consider a CP VLB within the one-photon regime. The spinors of the Hamiltonian in Eq.~\eqref{effectiveH} can be characterized by their total angular momentum, {\it e.g.}, for the $n^{\rm th}$-Floquet state $\mathcal{J}^{\rm F}_{z}\hat{\phi}_{n,j}(\bm r)= j \hat{\phi}_{n,j}(\bm r)$, where $j=\ell_e-(\ell +\tau+1)/2$, and $j\in \mathbb{Z}$ ($j\in\mathbb{Z}+1/2$) for $\ell$ even (for $\ell$ odd). Using this basis we obtain the radial differential equations that couple the different spinor components, 
\begin{equation}\label{reqns}
 \left[
   \begin{array}{cccc}
     -\widetilde{m}\footnotesize{+}\hbar\Omega & \mathscr{L}^{-}_{j_-} & 0 & \tilde{A}_{-} \\
     \mathscr{L}^{+}_{j_{-}-1} & \widetilde{m} & \tilde{A}^{*}_{+} & 0 \\
     0 & \tilde{A}_{+} & -\widetilde{m} &  \mathscr{L}^{-}_{j_+} \\
      \tilde{A}^{*}_{-}  & 0 & \mathscr{L}^{+}_{j_{+}-1} &  \widetilde{m}-\hbar\Omega \\
   \end{array}
 \right]\left[
          \begin{array}{c}
            \phi^{1}_{s} \\
            \phi^{1}_{\bar{s}} \\
            \phi^{0}_{s} \\
            \phi^{0}_{\bar{s}} \\
          \end{array}
        \right]=E^{n}_{j}\left[
          \begin{array}{c}
            \phi^{1}_{s} \\
            \phi^{1}_{\bar{s}} \\
            \phi^{0}_{s} \\
            \phi^{0}_{\bar{s}} \\
          \end{array}
        \right]
\end{equation}
where $\widetilde{m}=\hbar\Omega/2-M$, $\tilde{A}_{\pm}=\tilde{A}(r)(1\pm\tau )e^{i\theta(1\mp \tau)}/2$, $ \mathscr{L}^{\pm}_{\nu}=-i\hbar v_{\rm F}(\partial_r\mp \nu/r)$, $j_{\pm}=j\pm (\ell+\tau\pm 1)/2$, $s=(\uparrow,n,j)$ [$\bar{s}=(\downarrow,n,j)$], $E^{n}_{j}=\epsilon_{n,j}-\hbar\Omega/2$, and $\epsilon_{n,j}$ is the quasienergy of the $n^{\rm {th}}$ Floquet state with angular momentum $j$. 
\begin{figure}[ht!]
  \centering
  \includegraphics[width=0.5\textwidth]{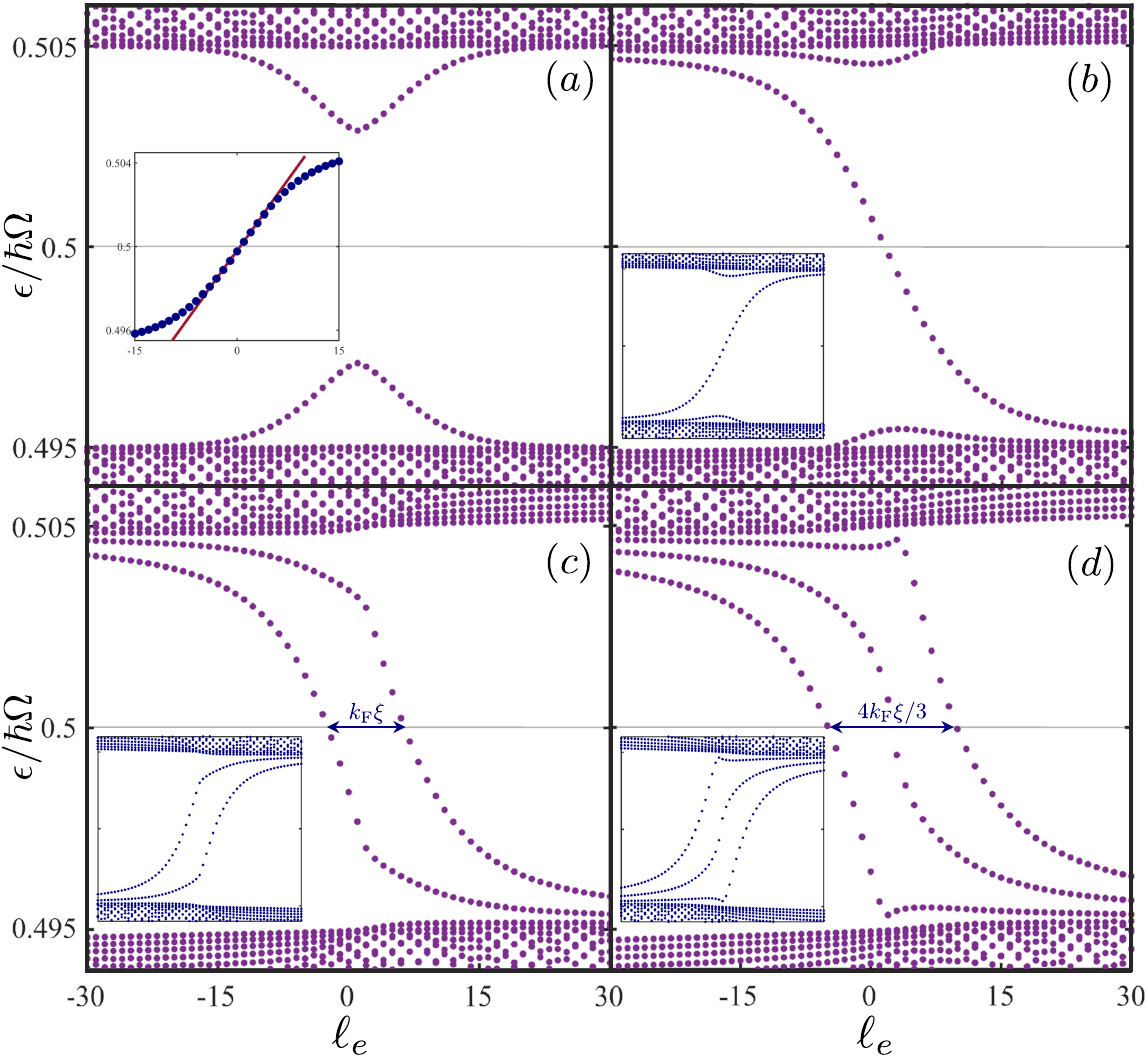}
  \caption{Quasienergy spectrum for a VLB-irradiated massive Dirac-like material with a gap $2M$. The driven system's parameters are $\hbar\Omega=2.06M$, $k_{\rm F}\xi=10$, $\Delta_0=0.01M$, $\Delta(r)=\Delta_0\tanh(r/\xi)$, $R=10\xi$, and $v_{\rm F}/\Omega=1$\AA. Panels (a)-(d) correspond to the spectra for VLBs with OAM $\ell=0$-$3$, respectively. The inset of (a) shows the CdGM solution for $\ell=-1$ which is shown in panel (b) inset. The solid red line shows the fitting to the analytical dispersion of CdGM states. Insets of (c) and (d) show solutions for OAM $\ell=-2$ and $-3$.}
  \label{Fig2}
\end{figure}

The diagonalization of the radially dependent two-band Floquet Hamiltonian is typically a numerical task. Fig.~\ref{Fig2} shows examples of quasienergy spectra as a function of the electron OAM $\ell_e$ obtained by numerically solving Eq.~\eqref{reqns} by implementing the Bessel decomposition method~\cite{B-decomp}. The same procedure can be applied to full Hamiltonian as shown in~\cite{supp}.  
However, deep insights could be gained by reducing the four linearly coupled differential equations to two coupled second order ones. To do so, we first recall that massive Dirac fermions couple strongly to a $\tau$-handed CP light while they couple weakly to $\bar{\tau}=-\tau$ due to their chiral nature~\cite{Dch1,Dch2,Dh3}. Hence, in the weak coupling regime, $g\ll1$ and the ratio between the $\bar{\tau}$ and $\tau$'s dynamical gap is $\widetilde{m}/\hbar\Omega$, making the $\bar{\tau}$ gap negligible ($\sim g\widetilde{m}$) for $\hbar\Omega\gtrapprox 2M$. Consequently, we can fix $\tau=1$, then apply $\mathscr{L}^{+}_{j_{-}-1}$ to the first equation and $\mathscr{L}^{-}_{j_+}$ to the fourth one. Additionally, for light frequencies close but larger than the semiconducting gap, $2\widetilde{m}/\hbar\Omega\ll1$, and for energies in the vicinity of the first photon resonance, $2E^{n}_{j}/\hbar\Omega \ll 1$, the coupled equations reduce to
\begin{equation}\label{mappedH}
- \frac{\hbar^2\sigma_z}{2m^{*}}\left[\partial_r^2+ \frac{\partial_r}{r} - \frac{\alpha_\ell^2}{r^2} + k^2_{\rm F} -\sigma_z\frac{\beta}{r^2}\right]\hat{\Psi} + \sigma_x\Delta(r)\hat{\Psi} = \mathscr{E}\hat{\Psi}.
\end{equation}
Strikingly, these radial equations can be mapped to the Bogoliubov–de Gennes equations (BdGEs) for an $s$-wave superconductor (SC) or superfluid with an $\ell$-flux-quanta vortex~\cite{multiply,Caroli,ScBook,SCMVE2}. Within this mapping $m^{*}=\hbar\Omega/(2v^2_{\rm F})$ plays the role of the effective quasielectron's mass, the Fermi wave vector $k_{\rm F}$  satisfies $\hbar^2k^2_{\rm F}/2m^{*}=\widetilde{m}$, and $\tilde{A}(r)$ corresponds to the magnitude of the superconducting order parameter with vorticity $\ell$, $\Delta(\bm r)=\Delta(r)e^{i\ell\theta}$. Then, the VLB's OAM, $\ell$ plays the role of the flux threading the SC, $\alpha_\ell=\sqrt{j^2+\ell^2}$, $\beta=j\ell$, and $\mathscr{E}=E^{n}_j$ is given in Eq.~\eqref{reqns}. Hence, physical properties of the VLB driven massive Dirac system, such as its bulk spectrum, can be directly inferred from the BdGEs governing a flux threaded SC with multiply quantized vortices~\cite{multiply}.       
       
In Fig.~\ref{Fig2}, we show the quasienergy spectrum obtained from the BdGE mapping, which coincides with the diagonalization of the effective Floquet Hamiltonian [Eq.~\eqref{reqns}] using Bessel decomposition. In order to gain insights into the nature and symmetries of the bands, we exploit the correspondence to superconductors further by analyzing spectra for different values of $\ell$. We begin with the general expression for the total angular momentum $j$ for $\tau=1$: $j=\ell_e-1-\ell/2$. If the light's OAM is $\ell=0$, in Fig.~\ref{Fig2} (a), the bulk of the irradiated system does not host vortex states at the first photon resonance as it mimics a SC with no vortices. We also note that due to the light's polarization the bulk spectrum satisfies $\mathscr{E}_{j+1}=\mathscr{E}_{-j+1}$. When the OAM of the VLB is $|\ell|=1$ as shown in Fig.~\ref{Fig2} (b) ($\ell=1$) and its inset ($\ell=-1$), Eq.~\eqref{mappedH} maps to the Caroli-de Gennes-Matricon (CdGM) model for vortex core states in $s$-wave SCs~\cite{CdGM}. Here a single vortex branch is displayed within the dynamical gap at $\hbar\Omega/2$. The vortex branch energy, $\mathscr{E}^{\rm V}_{\ell_e}$, exhibits the symmetry $\mathcal{S}_1:\mathscr{E}^{\rm V}_{j+1+\frac{\ell}{2}}=-\mathscr{E}^{\rm V}_{-j+1+\frac{\ell}{2}}$ which is analogous to the particle-hole symmetry displayed in the CdGM vortex core states. Moreover, for $\mathscr{E}\ll \widetilde{m}$ the vortex branch disperses linearly, $\mathscr{E}^{\rm V}_{\ell_e}=-\ell j\omega_0$, as shown in Fig.~\ref{Fig2} (a-inset). The states' separation is $\omega_0\sim\Delta_0/(k_{\rm F}\xi)$ where $\Delta_0=g\hbar\Omega/2$, $k_{\rm F}$ is given in Eq.~\eqref{mappedH}, and $\xi$ is the light-vortex saturation width [Fig.~\ref{Fig1} (a)].

Like $s$-wave fermionic superfluids or SCs with multiply quantized vortices with winding number $\ell$~\cite{multiply}, the VLB driven system with OAM $|\ell|> 1$ displays similar vortex states. In Fig.~\ref{Fig2} (b) (inset) $\ell=2$ ($\ell=-2$), two in-gap vortex states branches are displayed. For energies close to $\hbar{\Omega}/2$ ($\mathscr{E}\ll \widetilde{m}$) the two vortex branches disperse as $\mathscr{E}^{\rm {V}_{\pm}}_{\ell_e}\approx-\ell \omega_0 (j - \ell_{\pm})$, where $\ell_{\pm}\approx \pm k_{\rm F}\xi/2$. The maximum separation between the two vortex branches is $\ell_+-\ell_-\approx k_{\rm F}\xi$ at $\epsilon/\hbar\Omega \approx1/2$. Moreover, the values of $\ell_e$ and $\ell_{\pm}$ are constrained by $(2+\ell)/2<\ell_e<\pm\ell_{\pm}$, a range of values that defines the region where the vortex states branches disperse linearly. Beyond these values they begin to merge with the continuum. Unlike the case of $|\ell|=1$ the vortex branches for $|\ell|=2$ are not spectrally symmetric with respect to themselves ({\it i.e.} they do not exhibit symmetry $\mathcal{S}_1$). Instead they exhibit $\mathcal{S}_2:\mathscr{E}^{\rm {V}_{+}}_{j+1+\frac{\ell}{2}}=-\mathscr{E}^{\rm {V}_{-}}_{-j+1+\frac{\ell}{2}}$. For odd values of  VLB's OAM the system exhibits an odd number of vortex branches. Fig.~\ref{Fig2} (d), shows vortex branches for $|\ell|=3$. There are two outer branches $\mathscr{E}^{\rm {V}_{\pm}}_{\ell_e}$ and a central one $\mathscr{E}^{\rm {V}_{0}}_{\ell_e}$. For small quasienergies in the gap, the central branch disperses as $\mathscr{E}^{\rm {V}_{0}}_{\ell_e}\approx-\ell \omega_0 j$, while the outer branches' dispersion is $\mathscr{E}^{\rm {V}_{\pm}}_{\ell_e}\approx-\ell \omega_0 (j-\tilde{\ell}_{\pm})$ where $\tilde{\ell}_{\pm}\approx \pm 2k_{\rm F}\xi/3$. The separation between the outer branches is $\tilde{\ell}_{+}-\tilde{\ell}_{-}=4k_{\rm F}\xi/3$ and it is approximately twice their separation from the central one. Similar to the vortex branch for $|\ell|=1$ the central branch exhibits the $\mathcal{S}_1$ symmetry with $|\ell|=3$, while the two outer branches display $\mathcal{S}_2$ with $|\ell|=3$. 

These results can be generalized for a CP VLB carrying even or odd values of $|\ell|$. Systems driven with even values of $|\ell|$ will display $|\ell|/2$ pairs of vortex states branches within the dynamical gap generated at the edge of the Floquet Brillouin zone, $\hbar\Omega/2$. For quasienergies with $\mathscr{E}\ll \widetilde{m}$, these branches disperse linearly $\mathscr{E}^{\rm {V}_{q}}_{\ell_e}\approx-\omega_0\ell(j-\ell_{\pm q})$, where $0<q\le |\ell|/2$ represents the branch pair number, and $\ell_{\pm q}\approx\pm (2q-1)k_{\rm F}\xi/|\ell|$. The vortex states branches' separation is $\ell_{q}-\ell_{-q}$ and within a pair the branches are symmetric with respect to one another. On the other hand, for a VLB with odd $|\ell|$, in addition to the central vortex branch that linearly disperses as $\mathscr{E}^{\rm {V}_0}_{\ell_e}$, there are $(|\ell|-1)/2$ pairs of vortex branches that disperse as $\mathscr{E}^{\rm {V}_{q}}_{\ell_e}\approx-\omega_0\ell(j-\tilde{\ell}_{\pm q})$ with  $0<q\le (|\ell|-1)/2$, $\tilde{\ell}_{\pm q}\approx \pm 2qk_{\rm F}\xi/|\ell|$. The relative separation of the vortex branches in the pair $q$ is $\tilde{\ell}_{q}-\tilde{\ell}_{-q}$, and they satisfy the symmetry $\mathcal{S}_2$, while the central branch respects $\mathcal{S}_1$.  

In conclusion, we have presented a periodic-driving-based mechanism in Dirac-like materials that realizes multiply quantized vortex states when irradiated by VLBs. By means of Floquet theory, we found the conditions the VLB's parameters must satisfy for the driven system to host vortices characterized by a conserved total angular momentum. The emergence of these vortices is caused by the electron-hole hybridization due to photon absorption processes with the corresponding OAM transfer. Consequently, these vortices are not displayed in the off-resonant driving regime but only under resonant driving. Moreover, due to the VLB's polarization constraints, the conservation of total angular momentum implies that the irradiated system hosts vortex states exclusively when subjected to CP VLBs. We also show that in the weak light-matter coupling regime and for driving frequencies larger but comparable to the material's gap, the driven system's spectrum mimics that of fermionic superfluids and $s$-wave superconductors with multiply quantized vortices. We note that in addition to the vortex states generated by the CP VLB in the bulk of the massive Dirac material, the circular polarization carried by the VLB will also generate topological edge states~\cite{FloquetTI}, that will be discussed in future works. 

In Dirac materials with multiple valleys~\cite{Dch1}, the valley selective coupling to CP VLBs could obstruct the observation of the induced vortex states due to the vanishing gap at the weakly coupled valley. However, thin-films of topological insulators~\cite{Dh3,thin3,thin1} host a single valley of massive Dirac fermions at the $\Gamma$-point~\cite{thin2,thin4}, allowing to bypass this hurdle. Hence, this property positions them as plausible material candidates for the potential observation of light-induced vortices. We suggest that probes with atomic-spatial and femtosecond-temporal resolutions, such as time-resolved lightwave-driven scanning tunnelling spectroscopy~\cite{TimeSTM}, may be used to detect multiply quantized light-induced vortices. 

\begin{acknowledgements}
This work is supported by the NSF Grant No. DMR-2213429 (L.I.M, C.M, and M.M.A).
\end{acknowledgements}    

\bibliography{References.bib}
\newpage
\appendix
\renewcommand{\theequation}{S.\arabic{equation}}
\begin{widetext}

\section*{Supplementary Material}

\setcounter{section}{0}
\section{Floquet Hamiltonian of VLB Driven Massive Dirac-like Material}
Consider a $2$D massive Dirac-like material, with a gap $2M$, described by the Hamiltonian $H_{\rm D}(\bm p)=v_{\rm F} {\bm \sigma}\cdot\bm p+M\sigma_z$, where $\bm p=(p_x,p_y,0)$, and $\bm \sigma= (\sigma_x,\sigma_y,\sigma_z)$. When the material is subjected to a vortex light beam (VLB) with a polarization $\eta$ and orbital angular momentum $\ell$, we can capture the light's effects on the material's quasielectrons via minimal coupling, {\it i.e.} $p_i\rightarrow p_i+eA_i(\bm r, t)$ ($e>0$), which leads to the space-time-dependent Hamiltonian
\begin{equation}\label{t-Hamiltonain}
H({\bm r}, t)= v_{\rm F}{\bm \sigma}\cdot {\bm p}+M\sigma_z+ e v_{\rm F} {\bm{\mathcal{{A}}}}({\bm r, t}) \cdot {\bm\sigma}\;, 
\end{equation} 
where ${\bm{\mathcal{{A}}}}({\bm r, t})=A(r)\Re[e^{i(\Omega t -\ell\theta)}\varepsilon_{\eta}]$, $\theta$ is the azimuthal angle, $\varepsilon_{\eta}=\hat{x}+e^{-i\eta}\hat{y}$, and $\eta \in [0,2\pi)$. The time-periodic nature of the Hamiltonian makes it convenient to use Floquet's theory as it grants the existence of a set of solutions of the form  
\begin{equation}\label{states}
\hat{\psi}_n(\bm r, t) = e^{-i\epsilon_{n}t/\hbar}\hat{\phi}_n(\bm r, t)\;,\; {\rm where}\; \hat{\psi}^{\rm T}_n(\bm r, t)=[\psi_{n,\uparrow}({\bm r}, t), \psi_{n,\downarrow}({\bm r}, t)]\;,\; {\rm and}\;  \hat{\phi}^{\rm T}_n(\bm r, t)=[\phi_{n,\uparrow}({\bm r},t), \phi_{n,\downarrow}({\bm r},t)].
\end{equation} 
Here $\epsilon_n$ is the quasienergy modulo $\hbar\Omega$, and the Floquet states satisfy $\hat{\phi}^{\rm T}_n(\bm r, t+2\pi/\Omega)=\hat{\phi}^{\rm T}_n(\bm r, t)$. Substituting $\hat{\psi}_n(\bm r, t)$ in the time-dependent Schr\"{o}dinger equation, $H({\bm r}, t)\hat{\psi}_n(\bm r, t)=i\hbar\partial_t\hat{\psi}_n(\bm r, t)$ yields
\begin{equation}\label{FLSC}
\left[H({\bm r}, t) -i\hbar\partial_t\right]\hat{\phi}_n(\bm r, t)=\epsilon_n \hat{\phi}_n(\bm r, t)\;,
\end{equation} 
where $H_{\rm F}=H({\bm r}, t) -i\hbar\partial_t$ is the Floquet Hamiltonian. The time-periodicity of the Floquet states allows their representation as a Fourier series, 
\begin{equation}\label{FS}
\hat{\phi}_n(\bm{r},t)=\sum_{m=-\infty}^{\infty}e^{i m \Omega t} \hat{\phi}^{m}_{n}({\bm r})\;,  
\end{equation}
and by substituting these states in Eq.~\eqref{FLSC}, one arrives at 
\begin{equation}\label{FHgeneral}
\sum_{m}H^{\rm F}_{m'm}(\bm r)\hat{\phi}^{m}_n(\bm r)=\sum_{m}\left[H_{m'm}(\bm r) +m\hbar\Omega\delta_{m',m}\right]\hat{\phi}^{m}_n(\bm r)=\epsilon_{n}\hat{\phi}^{m'}_n(\bm r),
\end{equation}
where
\begin{equation}\label{FMatrix}
H_{m'm}(\bm r)=\frac{\Omega}{2\pi}\int_{0}^{2\pi/\Omega}H({\bm r}, t)e^{i(m-m')\Omega t}dt\;,
\end{equation}
is known as the Floquet matrix, and the matrix components of the Floquet Hamiltonian in the extended Hilbert space are $H^{\rm F}_{m'm}(\bm r)=H_{m'm}(\bm r)+m\hbar\Omega\delta_{m',m}$.  

With the definitions above and in the Floquet space basis, $\hat{\phi}^{\rm T}_n({\bm r})=[\dots, \hat{\phi}^{1}_n({\bm r}),\hat{\phi}^{0}_n({\bm r}),\hat{\phi}^{-1}_n({\bm r}),\dots ]$ we find the matrix elements of the Floquet Hamiltonian describing the VLB-driven massive Dirac material, such that 
\begin{subequations}\label{Hns}
\begin{eqnarray}
 && H^{\rm F}_{m'm}(\bm r) = \left[H_0(\bm r)+m\hbar\Omega\right]\delta_{m',m}+H_{+}(\bm r)\delta_{m',m+1} +H_{-}(\bm r)\delta_{m',m-1}, \;{\rm where} \\
 && H_0(\bm r)= v_{\rm F} {\bm \sigma}\cdot\bm p+M\sigma_z,  \\
 && H_{+}(\bm r) =  \frac{\tilde{A}(r)}{2}e^{-i\ell\theta}\left[ (1-ie^{-i\eta})\sigma_{+}+(1+ie^{-i\eta})\sigma_{-} \right],\; {\rm and} \\
 && H_{-}(\bm r) = \frac{\tilde{A}(r)}{2}e^{i\ell\theta}\left[ (1+ie^{i\eta})\sigma_{-}+(1-ie^{i\eta})\sigma_{+} \right], 
\end{eqnarray}
\end{subequations}
where $\tilde{A}(r)=ev_{\rm F}A(r)$ and these equations are equivalent to Eq.~(2) in the main text. 

\subsection{High-frequency (van Vleck) Approximation }
When the system a driven off-resonantly ($\hbar\Omega\gg W$ where $W$ is the material's bandwidth), irradiation effects are well-captured via the van Vleck approximation. This approximation uses the blocks of the Floquet Hamiltonian and to find the light-induced terms in powers of $\Omega^{-1}$. Within the van Vleck approximation and to the leading term in $\Omega^{-1}$~\cite{VanVleck}, we have
\begin{equation}\label{vVeffe}
H^{\rm v}_{\rm eff}(\bm r)\approx H_{0}(\bm r)+\frac{[H_{+}(\bm r),H_{-}(\bm r)]}{\hbar\Omega}+\mathcal{O}(\Omega^{-2}).
\end{equation}

For the VLB driven massive Dirac material, $H_{\pm }$ are given in Eq.~\eqref{Hns}, and 
\begin{equation}\label{commutationvV}
\left[H_+(\bm r),H_{-}(\rm r) \right]=\frac{\tilde{A}^2(r)}{4}\left\{(1-ie^{-i\eta})(1+ie^{i\eta})-(1-ie^{i\eta})(1+ie^{-i\eta})\right\}[\sigma_+,\sigma_-]=-\tilde{A}^2(r)\sin\eta\sigma_z\;,
\end{equation}
then the effective Hamiltonian in the high-frequency off-resonant regime is 
\begin{equation}\label{HvV}
H^{\rm v}_{\rm eff}(r)\approx v_{\rm F} {\bm \sigma}\cdot\bm p+\left(M-\frac{\tilde{A}^2(r)}{\hbar\Omega}\sin\eta\right)\sigma_z,
\end{equation}
clearly showing that the light-induced perturbation in this regime is OAM independent, as mentioned in the main text.

\section{Angular momentum operator and its conservation}
In this section, we find the total angular momentum operator that commutes with the full Floquet Hamiltonian and the constraints this imposes on the VLB's parameters. The Floquet Hamiltonian in the extended Hilbert space is given in Eq.~\eqref{Hns}, then we assume that there exists a total angular momentum operator in Floquet space, $\mathcal{J}_z^{\rm F}$, with matrix elements that take the form, 
\begin{equation}\label{Jz}
\mathcal{J}^{\rm F}_{z,m'm}=\left[\mathcal{L}_z+h(\eta,\ell)\frac{\sigma_z}{2}+f(\eta,\ell)\left(m-\frac{1}{2}\right)\sigma_0\right]\delta_{m',m},
\end{equation}   
where $h(\eta,\ell)$ and $f(\eta,\ell)$ are arbitrary functions of the light's polarization $\eta$ and the OAM $\ell$. We also recall the commutation relations, 
\begin{subequations}
\begin{eqnarray}
&&[\mathcal{L}_z, H_{0}(\bm r)] = iv_{\rm F}({\bm \sigma} \times {\bm p})_{\hat{z}}=-\left[\frac{\sigma_z}{2}, H_{0}(\bm r)\right],\; [\mathcal{L}_z, H_{\pm}(\bm r)] =\mp\ell H_{\pm}(\bm r),\\
&&\left[\frac{\sigma_z}{2}, H_{+}(\bm r)\right]=\frac{\tilde{A}(r)}{2}e^{-i\ell\theta}\left[ (1-ie^{-i\eta})\sigma_{+}-(1+ie^{-i\eta})\sigma_{-} \right],\; {\rm and}\\
&&\left[\frac{\sigma_z}{2}, H_{-}(\bm r)\right]=\frac{\tilde{A}(r)}{2}e^{i\ell\theta}\left[ -(1+ie^{i\eta})\sigma_{-}+(1-ie^{i\eta})\sigma_{+} \right].
\end{eqnarray}
\end{subequations}
In addition to these commutations, it is instructive to notice that 
\begin{subequations}
\begin{eqnarray}
  &&\left[\sigma_0\left(\mathbb{N}-\frac{\mathbb{I}}{2}\right), H^{\rm F}(\bm r)\right]_{m'm} = H_{+}(\bm r)\delta_{m',m+1} -H_{-}(\bm r)\delta_{m',m-1}, \\
  &&[\mathcal{L}_z\sigma_0\mathbb{I}, H^{\rm F}(\bm r)]_{m'm}= iv_{\rm F}({\bm \sigma} \times {\bm p})_{\hat{z}}\delta_{m',m}-\ell H_{+}(\bm r)\delta_{m',m+1}+\ell H_{-}(\bm r)\delta_{m',m-1}, \\
  &&\left[\frac{\sigma_z}{2}\mathbb{I}, H^{\rm F}(\bm r)\right]_{m'm}=-iv_{\rm F}({\bm \sigma} \times {\bm p})_{\hat{z}}\delta_{m',m}+\frac{\tilde{A}(r)}{2}e^{-i\ell\theta}\left[ (1-ie^{-i\eta})\sigma_{+}-(1+ie^{-i\eta})\sigma_{-} \right]\delta_{m',m+1}\nonumber\\
  &&{\;\;\;\;\;\;\;\;\;\;\;\;\;\;\;\;\;\;\;\;\;\;\;\;\;\;\;\;\;\;\;\;\;\;\;\;\;\;\;\;\;\;\;\;\;\;\;\;\;\;\;\;\;\;\;\;\;}+\frac{\tilde{A}(r)}{2}e^{i\ell\theta}\left[ -(1+ie^{i\eta})\sigma_{-}+(1-ie^{i\eta})\sigma_{+} \right]\delta_{m',m-1},
\end{eqnarray}
\end{subequations}
where $[A,B]_{mn}$ refers to the $mn$ element of the commutator of $A$ and $B$, $\mathbb{I}_{\rm F}$ is an identity matrix in Floquet space, $\mathbb{N}_{\rm F}={\rm diag}({\bm n})$, and $\bm n=(\dots,n,n-1,\dots,1,0,-1, \dots,-n+1,-n,\dots)$.

Then, assuming that $[\mathcal{J}_z^{\rm F},H^{\rm F}(\bm r)]=\mathcal{M}$, and using the relations above we find the elements of $\mathcal{M}$
\begin{eqnarray}\label{M}
\mathcal{M}_{m'm}=i({\bm \sigma} \times {\bm p})_{\hat{z}}\left[1-h(\eta,\ell)\right]\delta_{m'm}&&+\frac{\tilde{A}(r)}{2}e^{-i\ell\theta}\left\{ [f(\eta,\ell)+1-\ell][1-ie^{-i\eta}]\sigma_+ + [f(\eta,\ell)-1-\ell][1+ie^{-i\eta}]\sigma_-\right\}\delta_{m',m+1}\nonumber\\
&&-\frac{\tilde{A}(r)}{2}e^{i\ell\theta}\left\{[f(\eta,\ell)+1-\ell][1+ie^{i\eta}]\sigma_{-} +[f(\eta,\ell)-1-\ell][1-ie^{i\eta}]\sigma_{+} \right\}\delta_{m',m-1}\nonumber\;.\\
\end{eqnarray}
Requiring $[\mathcal{J}_z^{\rm F},H^{\rm F}]=0$ implies that all the elements of $\mathcal{M}$ must be zero. Hence for the diagonal elements to vanish $h(\eta,\ell)=1$, moreover, the off-diagonal elements become zero if the following four equations are simultaneously satisfied,
\begin{equation}
[f(\eta,\ell)+1-\ell][1-ie^{-i\eta}] = 0,\;[f(\eta,\ell)-1-\ell][1+ie^{-i\eta}] = 0,\; [f(\eta,\ell)+1-\ell][1+ie^{i\eta}] = 0,\; [f(\eta,\ell)-1-\ell][1-ie^{i\eta}]  = 0\;.
\end{equation}
These equations are concomitantly satisfied in the following two cases: (1) $\eta=-\pi/2$ then $f(\eta,\ell)=\ell-1$, and (2) $\eta=\pi/2$ then $f(\eta,\ell)=\ell+1$. On the other hand, if $\eta\ne \pm\pi/2$ we arrive at the condition $f(\eta,\ell)=\ell-1\;{\rm and}\;f(\eta,\ell)=\ell+1 $ simultaneously. Hence, $\mathcal{J}_z^{\rm F}$ is a symmetry of the non-perturbative Floquet {\it if and only if} the VLB is circularly polarized, therefore the total angular momentum operator takes the form  

\begin{equation}\label{Jz}
\mathcal{J}_z^{\rm F}=\left(\mathcal{L}_z\sigma_0+\frac{\sigma_z}{2}\right)\mathbb{I}_{\rm F}+(\ell+\tau)\sigma_0\left(\mathbb{N}_{\rm F}-\frac{\mathbb{I}_{\rm F}}{2}\right),
\end{equation}
where $\tau=\pm1$ determines the circular polarization's handedness.  
\subsection{Projected Floquet Hamiltonian and Total Angular Momentum Operator} 
To obtain the effective Floquet Hamiltonian we consider the values of $m'=\{1,0\}$ in Eq.~\eqref{Hns}, which can be written in a matrix form as 
\begin{equation}\label{MatixHf}
H^{\rm F}_{\rm eff}(\bm r)=\left(
                             \begin{array}{cc}
                               v_{\rm F} {\bm \sigma}\cdot\bm p+M\sigma_z+\hbar\Omega\sigma_0 & H_{+}(\bm r) \\
                               H_{-}(\bm r) & v_{\rm F} {\bm \sigma}\cdot\bm p+M\sigma_z \\
                             \end{array}
                           \right)=\left[H_0(\bm r)+\frac{\hbar\Omega}{2}\sigma_0\right]\alpha_{0}+\frac{\hbar\Omega}{2}\sigma_0\alpha_z+H_{+}(\bm r)\alpha_+ +H_{-}(\bm r)\alpha_- \;.
\end{equation}
Similarly, the total angular momentum in this subspace, Eq.~\eqref{Jz}, becomes 
\begin{equation}\label{JZproj}
\mathcal{J}_z^{\rm F}=\left(
  \begin{array}{cc}
    \mathcal{L}_z+\sigma_z/2+(\ell+\tau)/2 & 0 \\
    0 & \mathcal{L}_z+\sigma_z/2-(\ell+\tau)/2 \\
  \end{array}
\right)=\left(\mathcal{L}_z\sigma_0+\frac{\sigma_z}{2}\right)\alpha_0+(\ell+\tau)\frac{\sigma_0\alpha_z}{2}\;.
\end{equation} 
Here the $\alpha$ matrices are the Pauli matrices acting on the reduced Floquet space.   
\subsection{Eigenstates and Radial Equations of the Floquet Hamiltonian}
For circularly polarized VLBs, $\mathcal{J}_z^{\rm F}$ [Eq.~\eqref{Jz}] is a symmetry of the Floquet Hamiltonian [Eq.~\eqref{Hns}], then the Floquet states in Eq.~\eqref{FHgeneral} are characterized by their total angular momentum, such that, $\mathcal{J}_z^{\rm F}\hat{\phi}_{n,j}(\bm r)= j \hat{\phi}_{n,j}(\bm r)$. Based on this symmetry we can construct the $\hat{\phi}_{n,j}(\bm r)$ states, such that 
\begin{subequations}\label{Phij}
\begin{eqnarray}
&&\hat{\phi}^{\rm T}_{n,j}(\bm r)=\left[\dots \hat{\phi}^{m, \rm T}_{n,j}(\bm r), \dots,\hat{\phi}^{0, \rm T}_{n,j}(\bm r),\dots,\hat{\phi}^{-m, \rm T}_{n,j}(\bm r),\dots\right],\;\\
&&\hat{\phi}^{m, \rm T}_{n,j}(\bm r)= e^{-i(\ell+\tau)m\theta}e^{i\ell_e\theta}\hat{\varphi}^{m, \rm T}_{n,j}(\bm r), \; {\rm and}\\
&&\hat{\varphi}^{m, \rm T}_{n,j}(\bm r)= \left[ \phi^{m}_{n,\uparrow,j}(r)e^{-i\theta}, \phi^{m}_{n,\downarrow,j}(r) \right]\;,
\end{eqnarray}
\end{subequations}
where the $\hat{\phi}(\bm r)$ represent spinors, and $\phi^{m}_{n,s,j}(r)$ [$s=(\uparrow,\downarrow)$] are the radial components of the Floquet space spinor. When we apply $\mathcal{J}_z^{\rm F}$ in Eq.~\eqref{Jz} to this spinor we get, 
\begin{equation}\label{JphiValue}
\mathcal{J}_z^{\rm F}\hat{\phi}_{n,j}(\bm r)=\left[\ell_e-\frac{(\ell+\tau+1)}{2}\right]\hat{\phi}_{n,j}(\bm r)=j\hat{\phi}_{n,j}\;,
\end{equation} 
where $j\in \mathbb{Z}$ ($j\in\mathbb{Z}+1/2$) for $\ell$ even (for $\ell$ odd).     

It is useful to notice that
\begin{subequations}\label{props}
\begin{eqnarray}
 && H_{+}(\bm r)\hat{\phi}^{m-1}_{n,j}(\bm r) = e^{i\tau\theta}F_{+}(\tau)e^{-i(\ell+\tau)m\theta}e^{i\ell_e\theta}\hat{\phi}^{m-1}_{n,j}(\bm r)\;, \\
 && H_{-}(\bm r)\hat{\phi}^{m+1}_{n,j}(\bm r) = e^{-i\tau\theta}(\tau)F_{-}(\tau)e^{-i(\ell+\tau)m\theta}e^{i\ell_e\theta}\hat{\phi}^{m+1}_{n,j}(\bm r)\;,\\
 && F_{\pm}(\tau)=\frac{\tilde{A}(r)}{2}[(1+\tau)\sigma_{\mp}+(1-\tau)\sigma_{\pm}]\;.
\end{eqnarray}
\end{subequations}
The coupled equation governing $\hat{\phi}^{m}_{n,j}(\bm r)$ is 
\begin{equation}\label{mth}
H_{+}(\bm r) \hat{\phi}^{m-1}_{n,j}(\bm r) + H_{-}(\bm r) \hat{\phi}^{m+1}_{n,j}(\bm r) + [H_{0}(\bm r) + m\hbar\Omega ] \hat{\phi}^{m}_{n,j}(\bm r)= \epsilon_{n,j}\hat{\phi}^{m}_{n,j}(\bm r)\;.
\end{equation}
Using the properties in Eqs.~\eqref{props} and projecting on real space with polar coordinates, one arrives at the coupled differential equations 
\begin{eqnarray}
\frac{\tilde{A}(r)}{2}e^{i\tau\theta}\left[
  \begin{array}{c}
    (1-\tau)\phi^{m-1}_{n,\downarrow,j}(r) \\
    (1+\tau)\phi^{m-1}_{n,\uparrow,j}(r)e^{-i\theta}\\
  \end{array}
\right]+\frac{\tilde{A}(r)}{2}e^{-i\tau\theta}\left[
  \begin{array}{c}
    (1+\tau)\phi^{m+1}_{n,\downarrow,j}(r) \\
    (1-\tau)\phi^{m+1}_{n,\uparrow,j}(r)e^{-i\theta}\\
  \end{array}
\right]+\left[
          \begin{array}{c}
            (M+m\hbar\Omega)\phi^{m}_{n,\uparrow,j}(r)e^{-i\theta} \\
            (-M+m\hbar\Omega)\phi^{m}_{n,\downarrow,j}(r) \\
          \end{array}
        \right]\nonumber\\
+\left[
          \begin{array}{c}
           \mathscr{L}^{-}_{\ell_e-m(\ell+\tau)} \phi^{m}_{n,\downarrow,j}(r)e^{-i\theta} \\
            \mathscr{L}^{+}_{\ell_e-m(\ell+\tau)-1}\phi^{m}_{n,\uparrow,j}(r) \\
          \end{array}
        \right]=\epsilon_{n,j}\left[
          \begin{array}{c}
           \phi^{m}_{n,\uparrow,j}(r)e^{-i\theta} \\
            \phi^{m}_{n,\downarrow,j}(r) \\
          \end{array}
        \right]\;,
\end{eqnarray}
where $\mathscr{L}^{\pm}_{\nu}=-i\hbar v_{\rm F}(\partial_r\mp \nu/r)$ are differential operators. Then the coupled differential equations for the $m^{\rm th}$ Floquet component are
\begin{subequations}\label{coupledEqs}
\begin{eqnarray}
&&\mathscr{L}^{-}_{j-(m-\frac{1}{2})(\ell+\tau)+\frac{1}{2}} \phi^{m}_{n,\downarrow,j}(r)+ \left[M+\left(m-\frac{1}{2}\right)\hbar\Omega\right]\phi^{m}_{n,\uparrow,j}(r)\nonumber\\
&&+\frac{\tilde{A}(r)}{2}\left[ e^{i(\tau+1)\theta}(1-\tau)\phi^{m-1}_{n,\downarrow,j}(r)+ e^{-i(\tau-1)\theta}(1+\tau)\phi^{m+1}_{n,\downarrow,j}(r)\right]
=E^{n}_{j}\phi^{m}_{n,\uparrow,j}(r)\label{r1}\\
&&\mathscr{L}^{+}_{j-(m-\frac{1}{2})(\ell+\tau)-\frac{1}{2}}\phi^{m}_{n,\uparrow,j}(r)+\left[-M+\left(m-\frac{1}{2}\right)\hbar\Omega\right]\phi^{m}_{n,\downarrow,j}(r)\nonumber\\
&&+\frac{\tilde{A}(r)}{2}\left[e^{i(\tau-1)\theta}(1+\tau)\phi^{m-1}_{n,\uparrow,j}(r)+e^{-i(\tau+1)\theta}(1-\tau)\phi^{m+1}_{n,\uparrow,j}(r) \right]   
=E^{n}_{j}\phi^{m}_{n,\downarrow,j}(r)\;,\label{r2}
\end{eqnarray}
\end{subequations}
where $E^{n}_{j}=\left(\epsilon_{n,j}-\hbar\Omega/2\right)$. When the circularly polarized light is characterized by $\tau=1$, these equations reduce to
\begin{subequations}\label{coupledEqstau1}
\begin{eqnarray}
\mathscr{L}^{-}_{j-(m-\frac{1}{2})\ell-(m-1)} \phi^{m}_{n,\downarrow,j}(r)+ \left[M+\left(m-\frac{1}{2}\right)\hbar\Omega\right]\phi^{m}_{n,\uparrow,j}(r)+\tilde{A}(r) \phi^{m+1}_{n,\downarrow,j}(r)=E^{n}_{j}\phi^{m}_{n,\uparrow,j}(r)\\
\mathscr{L}^{+}_{j-(m-\frac{1}{2})\ell-m}\phi^{m}_{n,\uparrow,j}(r)+\left[-M+\left(m-\frac{1}{2}\right)\hbar\Omega\right]\phi^{m}_{n,\downarrow,j}(r)+\tilde{A}(r)\phi^{m-1}_{n,\uparrow,j}(r)= E^{n}_{j}\phi^{m}_{n,\downarrow,j}(r)\;.
\end{eqnarray}
\end{subequations}
\subsection{Eigenstates and Radial Equations of the Projected Floquet Hamiltonian}
Similar to the full Floquet Hamiltonian, the projected one in Eq.~\eqref{MatixHf} also conserves $\mathcal{J}_z^{\rm F}$ [Eq.~\eqref{JZproj}] uniquely for circularly polarized VLBs. Based on this conservation the states of the Hamiltonian are
\begin{equation}\label{effective_states}
\hat{\phi}^{\rm T}_{n,j}(\bm r)=\left[e^{-i(\ell+\tau+1)\theta}\phi^{1}_{n,\uparrow,j}(r), e^{-i(\ell+\tau)\theta}\phi^{1}_{n,\downarrow,j}(r), e^{-i\theta}\phi^{0}_{n,\uparrow,j}(r), \phi^{0}_{n,\downarrow,j}(r)\right]e^{i\ell_e\theta}\;.
\end{equation} 
When we apply $\mathcal{J}_z^{\rm F}$ in Eq.~\eqref{JZproj} to this state, similar to Eq.~\eqref{JphiValue}, we find
\begin{equation}\label{JphiValueEffective}
\mathcal{J}_z^{\rm F}\hat{\phi}_{n,j}(\bm r)=\left[\ell_e-\frac{(\ell+\tau+1)}{2}\right]\hat{\phi}_{n,j}(\bm r)=j\hat{\phi}_{n,j}\;.
\end{equation} 
The radial coupled differential equations of the projected Floquet Hamiltonian are obtained by considering the $m=\{1,0\}$ subspace in Eqs.~\eqref{r1} and \eqref{r2} to get
\begin{subequations}\label{EffectiveR}
\begin{eqnarray}
\mathscr{L}^{-}_{j-\frac{\ell+\tau-1}{2}} \phi^{1}_{n,\downarrow,j}(r)+ \left[M+\frac{\hbar\Omega}{2}\right]\phi^{1}_{n,\uparrow,j}(r)+\tilde{A}(r) e^{i(\tau+1)\theta}\frac{(1-\tau)}{2}\phi^{0}_{n,\downarrow,j}(r)
=E^{n}_{j}\phi^{1}_{n,\uparrow,j}(r)\;,\\
\mathscr{L}^{+}_{j-\frac{\ell+\tau-1}{2}-1}\phi^{1}_{n,\uparrow,j}(r)
-\left[M-\frac{\hbar\Omega}{2}\right]\phi^{1}_{n,\downarrow,j}(r)
+\tilde{A}(r)e^{i(\tau-1)\theta}\frac{(1+\tau)}{2}\phi^{0}_{n,\uparrow,j}(r)
=E^{n}_{j}\phi^{1}_{n,\downarrow,j}(r)\;,\\
\mathscr{L}^{-}_{j+\frac{\ell+\tau+1}{2}} \phi^{0}_{n,\downarrow,j}(r)+ \left[M-\frac{\hbar\Omega}{2}\right]\phi^{0}_{n,\uparrow,j}(r)
+\tilde{A}(r)e^{-i(\tau-1)\theta}\frac{(1+\tau)}{2}\phi^{1}_{n,\downarrow,j}(r)
=E^{n}_{j}\phi^{0}_{n,\uparrow,j}(r)\;,\\
\mathscr{L}^{+}_{j+\frac{\ell+\tau+1}{2}-1}\phi^{0}_{n,\uparrow,j}(r)
-\left[M+\frac{\hbar\Omega}{2}\right]\phi^{0}_{n,\downarrow,j}(r)
+\tilde{A}(r)e^{-i(\tau+1)\theta}\frac{(1-\tau)}{2}\phi^{1}_{n,\uparrow,j}(r)   
=E^{n}_{j}\phi^{0}_{n,\downarrow,j}(r)\;.
\end{eqnarray}
\end{subequations}
These equations are equivalent to Eq.~(5) in the main text. Moreover, when $\tau=1$ these equations become~\cite{Maria}
\begin{subequations}\label{es}
\begin{eqnarray}
&&\mathscr{L}^{-}_{j-\frac{\ell}{2}} \phi^{1}_{n,\downarrow,j}(r)+ \left[M+\frac{\hbar\Omega}{2}\right]\phi^{1}_{n,\uparrow,j}(r)
=E^{n}_{j}\phi^{1}_{n,\uparrow,j}(r)\;,\label{e1}\\
&&\mathscr{L}^{+}_{j-\frac{\ell}{2}-1}\phi^{1}_{n,\uparrow,j}(r)
-\left[M-\frac{\hbar\Omega}{2}\right]\phi^{1}_{n,\downarrow,j}(r)
+\tilde{A}(r)\phi^{0}_{n,\uparrow,j}(r)
=E^{n}_{j}\phi^{1}_{n,\downarrow,j}(r)\;,\label{e2}\\
&&\mathscr{L}^{-}_{j+\frac{\ell}{2}+1} \phi^{0}_{n,\downarrow,j}(r)+ \left[M-\frac{\hbar\Omega}{2}\right]\phi^{0}_{n,\uparrow,j}(r)
+\tilde{A}(r)\phi^{1}_{n,\downarrow,j}(r)
=E^{n}_{j}\phi^{0}_{n,\uparrow,j}(r)\;,\label{e3}\\
&&\mathscr{L}^{+}_{j+\frac{\ell}{2}}\phi^{0}_{n,\uparrow,j}(r)
-\left[M+\frac{\hbar\Omega}{2}\right]\phi^{0}_{n,\downarrow,j}(r)  
=E^{n}_{j}\phi^{0}_{n,\downarrow,j}(r)\;.\label{e4}
\end{eqnarray}
\end{subequations}
These equations are equivalent to the matrix equation, Eq.~(5), in the main text.   
\subsection{Mapping the VLB Driven System Equations to BdGEs}
Starting from the set of Eqs.~\eqref{es}, we apply the differential operator $\mathscr{L}^{+}_{j-\frac{\ell}{2}-1}$ to Eq.~\eqref{e1} and $\mathscr{L}^{-}_{j+\frac{\ell}{2}+1}$ to Eq.~\eqref{e4}, an using the properties  
\begin{equation}\label{properties}
\mathscr{L}^{+}_{\nu} \mathscr{L}^{-}_{\nu+1}=-(\hbar v_{\rm F})^2\left[\frac{\partial^2}{\partial r^2}+\frac{1}{r}\frac{\partial}{\partial r}-\frac{(\nu+1)^2}{r^2}\right],\; {\rm and},\; \mathscr{L}^{-}_{\nu+1} \mathscr{L}^{+}_{\nu}=-(\hbar v_{\rm F})^2\left[\frac{\partial^2}{\partial r^2}+\frac{1}{r}\frac{\partial}{\partial r}-\frac{\nu^2}{r^2}\right]\;,
\end{equation}
we find 
\begin{subequations}\label{mapping1}
\begin{eqnarray}
 \left[\frac{\partial^2}{\partial r^2}+\frac{1}{r}\frac{\partial}{\partial r}-\frac{(j-\frac{\ell}{2})^2}{r^2}+k^2_{1}\right]\phi^{1}_{n,\downarrow,j}(r)  &=& \frac{k^2_1\tilde{A}(r)}{(E^{n}_{j}-\widetilde{m})}\phi^{0}_{n,\uparrow,j}(r)\;,  \\
  \left[\frac{\partial^2}{\partial r^2}+\frac{1}{r}\frac{\partial}{\partial r}-\frac{(j+\frac{\ell}{2})^2}{r^2}+k^2_{0}\right]\phi^{0}_{n,\uparrow,j}(r)  &=& 
  \frac{k^2_0\tilde{A}(r)}{(E^{n}_{j}+\widetilde{m})}\phi^{1}_{n,\downarrow,j}(r)\;,
\end{eqnarray}
where $\widetilde{m}=\hbar\Omega/2-M$, 
\begin{equation}\label{ks}
 (\hbar v_{\rm F} k_1)^2=\left(E^{n}_{j}-\frac{\hbar\Omega}{2}\right)^2-\left(\frac{\hbar\Omega}{2}-\widetilde{m}\right)^2,\; {\rm and}\; (\hbar v_{\rm F} k_0)^2=\left(E^{n}_{j}+\frac{\hbar\Omega}{2}\right)^2-\left(\frac{\hbar\Omega}{2}-\widetilde{m}\right)^2\;. 
\end{equation}
\end{subequations}

In the regime where the light frequency is close, but larger than the semiconducting gap, $2\widetilde{m}/\hbar\Omega\ll1$, and for energies in the vicinity of the first photon resonance, $2E^{n}_{j}/\hbar\Omega \ll 1$, we have 
\begin{equation}\label{approx1}
(\hbar v_{\rm F} k_1)^2\approx\hbar\Omega(\widetilde{m}-E^{n}_{j}),\; (\hbar v_{\rm F} k_0)^2\approx\hbar\Omega(\widetilde{m}+E^{n}_{j}),\;{\rm and}\; \frac{k^2_0}{(E^{n}_{j}+\widetilde{m})}=-\frac{k^2_1}{(E^{n}_{j}-\widetilde{m})}\approx\frac{\hbar\Omega}{(\hbar v_{\rm F})^2}\;.
\end{equation}   
Then by setting $E^{n}_{j}\equiv\mathscr{E}$ Eq.~\eqref{mapping1} becomes
\begin{subequations}\label{mapping2}
\begin{eqnarray}
\frac{(\hbar v_{\rm F})^2}{\hbar\Omega}\left[\frac{\partial^2}{\partial r^2}+\frac{1}{r}\frac{\partial}{\partial r}-\frac{j^2+\frac{\ell^2}{4}}{r^2}+\frac{\hbar\Omega\widetilde{m}}{(\hbar v_{\rm F})^2}\right]\phi^{1}_{n,\downarrow,j}(r)+\tilde{A}(r)\phi^{0}_{n,\uparrow,j}(r)=\left[\mathscr{E}-\frac{(\hbar v_{\rm F})^2}{\hbar\Omega}\frac{\ell j}{r^2}\right]\phi^{1}_{n,\downarrow,j}(r)\;,\\
-\frac{(\hbar v_{\rm F})^2}{\hbar\Omega}\left[\frac{\partial^2}{\partial r^2}+\frac{1}{r}\frac{\partial}{\partial r}-\frac{j^2+\frac{\ell^2}{4}}{r^2}+\frac{\hbar\Omega\widetilde{m}}{(\hbar v_{\rm F})^2}\right]\phi^{0}_{n,\uparrow,j}(r)+\tilde{A}(r)\phi^{1}_{n,\downarrow,j}(r)=\left[\mathscr{E}-\frac{(\hbar v_{\rm F})^2}{\hbar\Omega}\frac{\ell j}{r^2}\right]\phi^{0}_{n,\uparrow,j}(r)\;.
\end{eqnarray}
\end{subequations}
In analogy with a superconductor (SC) or superfluid with an $\ell$-flux-quanta vortex~\cite{multiply,Caroli,ScBook}, we define 
\begin{equation}
\hat{\Psi}(r) = \left[
                 \begin{array}{c}
                   \phi^{0}_{n,\uparrow,j}(r) \\
                   \phi^{1}_{n,\downarrow,j}(r) \\
                 \end{array}
               \right],\; \alpha_\ell=\sqrt{j^2+\frac{\ell^2}{4}},\; \beta=\ell j,\; m^{*}=\frac{\hbar\Omega}{2v^2_{\rm F}},\;\frac{\hbar^2}{2m^{*}}k^2_{\rm F}= \widetilde{m},\;{\rm and}\; \Delta(r)=\tilde{A}(r).
\end{equation}
Therefore, Eq.~\eqref{mapping2} becomes 
\begin{equation}\label{mappedH}
- \frac{\hbar^2\sigma_z}{2m^{*}}\left[\partial_r^2+ \frac{\partial_r}{r} - \frac{\alpha_\ell^2}{r^2} + k^2_{\rm F} -\sigma_z\frac{\beta}{r^2}\right]\hat{\Psi}(r) + \sigma_x\Delta(r)\hat{\Psi}(r) = \mathscr{E}\hat{\Psi}(r).
\end{equation}

The equation above captures the bulk properties of the VLB irradiated system and has a one to one correspondence with the BdGEs governing an $s$-wave SC or superfluid hosting an $\ell$-flux-quanta vortex. Hence, following the procedure in Ref.~\cite{multiply} which relies on the small $r$ limit and asymptotic behaviors of the wave functions of Eq.~\eqref{mappedH}, one obtains an the vortex states energies near the one-photon resonance, as
\begin{equation}\label{vortex energies}
\mathscr{E^{\rm V}}= -\omega_0\ell j +\left(n+\frac{|\ell| -1}{2}\right)\tilde{\omega}\;,
\end{equation} 
where 
\begin{equation}\label{omegas}
\omega_0=\frac{2\tilde{ \omega}}{\pi k^{2}_{\rm F}}\int_{0}^{\infty}\frac{\Delta(r')}{r'}e^{-\frac{2}{k_{\rm F}}\int_{0}^{r'}\Delta(r'')\,dr''}\,dr' \; {\rm and}\;\tilde{\omega}=\frac{\pi}{2}\frac{k_{\rm F}}{\int_{0}^{\infty}e^{-\frac{2}{k_{\rm F}}\int_{0}^{r'}\Delta(r'')\,dr''}\,dr'}\;.
\end{equation}     
When the radial dependence is $\Delta(r)=\Delta_0\tanh(r/\xi)$ the integral above can be performed and the vortex states energies in Eq.~\eqref{vortex energies} become
\begin{equation}\label{omegas}
\mathscr{E^{\rm V}}=-\left(a \frac{\Delta_0}{k_{\rm F}\xi}\right)\ell j+b\left(n+\frac{|\ell| -1}{2}\right)\frac{k_{\rm F}}{\xi}\;,
\end{equation}  
where $b/a\equiv\gamma=\pi^3/[14\zeta(3)]\approx1.8$~\cite{multiply}. The vortex states with energies closest to the first photon resonance, $\mathscr{E^{\rm V}}\approx 0$, satisfy, 
\begin{equation}\label{jns}
j_{n}=\gamma\left(n+\frac{|\ell| -1}{2}\right)\frac{k_{\rm F}\xi}{\ell}\;,
\end{equation}    
and by imposing $|j|\lesssim k_{\rm F}\xi $ one gets $-|\ell|+1/2\lesssim n\lesssim 1/2$, which implies that $n$ can take the values of $0,-1,-2,-3, \dots 1-|\ell|$, demonstrating the existence of $\ell$ vortex states branches. For instance, when $|\ell|=1$, $n=0$, and we have a single vortex with energy $\mathscr{E}^{\rm V}_{\ell_e}=-\ell j \omega_0$, where $\omega_0\approx \Delta_0/(k_{\rm F}\xi)$. The relation between $\ell_e$ and $j$ is given in Eq.~\eqref{JphiValue}, and these states are equivalent to the CdGM vortex core states~\cite{CdGM}. When $\ell=2$, $n=0,-1$ which leads to $\mathscr{E}^{\rm V_{\pm}}_{\ell_e}=-\ell\omega_0(j\pm\ell_{\pm})$, where $\ell_+=-\ell_-$, and $\ell_+-\ell_-\approx\tilde{\omega}/(2 \omega_0)=k_{\rm F}\xi $. For $|\ell|=3$, $n=0,-1,-2$ and we have three vortex branches. The vortex branch associated with $n=-1$ disperses as $\mathscr{E}^{\rm V_0}_{j}=-\ell j \omega_0$, while the pair of vortex branches associated with $n=0$ and $-2$ disperse as $\mathscr{E}^{\rm V_\pm}_{\ell_e}=-\ell\omega_0(j\pm\ell_{\pm})$ where in this case $\ell_+-\ell_-\approx 4k_{\rm F}\xi/3$ and $\ell_+=-\ell_-$. Therefore, in general, when $0<|\ell|$ is an even number, we find that $\mathscr{E}^{\rm {V}_{q}}_{\ell_e}\approx-\omega_0\ell(j-\ell_{\pm q})$, where $\ell_{\pm q}\approx\pm (2q-1)k_{\rm F}\xi/|\ell|$, and $0<q\le |\ell|/2$ represents the branch pair number. For $|\ell|$ odd the central branch disperses as $\mathscr{E}^{\rm {V}_0}_{\ell_e}$, while the vortex pairs disperse as $\mathscr{E}^{\rm {V}_{q}}_{\ell_e}\approx-\omega_0\ell(j-\tilde{\ell}_{\pm q})$ with  $\tilde{\ell}_{\pm q}\approx \pm 2qk_{\rm F}\xi/|\ell|$ and $0<q\le (|\ell|-1)/2$.

\section{Numerical Solutions of the Floquet Hamiltonian's Radial Equations}
The diagonalization of the angular part of the Floquet Hamiltonian reduced the problem to a radial dependent problem. In order to find the quasienergy spectrum and the Floquet states, we numerically diagonalize the radial coupled equations shown in Eq.~\eqref{coupledEqs}. We adopt the Bessel decomposition technique (Fourier–Bessel series)~\cite{B-decomp}, in which one projects the radial wavefunctions onto the set of Bessel functions normalized in a disk of radius $R$. Thus, it is possible to diagonalize for each angular momentum $j$ separately. For numerical efficiency, we have also applied a $z$-rotation by $\pi/2$ to the spin basis of the Floquet Hamiltonian through the unitary operator $U=e^{-i\sigma_z\pi/4}$ that acts on the spin space of the Floquet spinor. In this basis, the wave functions for the $m^{\rm th}$ component of the spinor in Eq.~\eqref{Phij} are expressed as 
\begin{equation}\label{phims}
\hat{\Phi}^{m}_{n,j}(r)=U\hat{\phi}^{m}_{n,j}(r)=\sum_{\nu=1}^{N}\left[
    \begin{array}{c}
    a^{m}_{n,j,\nu}e^{-i\pi/4}\phi^{m}_{\nu,n,\uparrow,\mu_m}(r) \\
    b^{m}_{n,j,\nu}e^{i\pi/4}\phi^{m}_{\nu,n,\downarrow,\mu_m+1}(r) \\
    \end{array}
\right]
\end{equation}
where $\mu_m=j-(m-\frac{1}{2})(\ell+\tau)-\frac{1}{2}$,
\begin{equation}\label{bessels1}
\phi^{m}_{\nu,n,s,\mu}(r)=\frac{\sqrt{2}}{RJ_{\mu+1}(\alpha_{\nu,\mu})}J_{\mu}\left(\alpha_{\nu,\mu}\frac{r}{R}\right),\; \int_{0}^{R}\phi^{m}_{\nu,n,s,\mu}(r)\phi^{m}_{\nu',n,s,\mu}(r)rdr=\delta_{\nu,\nu'},
\end{equation}
and $\alpha_{\nu,\mu}$ is the $\nu^{\rm th}$ zero of the $\mu^{\rm th}$ Bessel function. Ideally the sum in Eq.~\eqref{phims} should extend to infinity as there is an infinite number of zeros of the Bessel functions, similarly the Floquet Hamiltonian contains an infinite number of components $m$. However, numerically we chose $N$ to be the cutoff of $\nu$ and $m_{\rm max}=(N_{\rm F}-1)/2$ ($N_{\rm F}$ is odd) to be the maximum Floquet component considered, {\it i.e.} $(m_{\rm max},\dots, 1,0,-1,\dots,-m_{\rm max})$. With these choices the size of the Floquet Hamiltonian matrix that is diagonalized at each $j$ is $N_{\rm F}(2N\times 2N)$ where the factors of $2$ are due to the spin, and this matrix equation takes the form    

\begin{eqnarray}\label{HmatL}
\left[
  \begin{array}{cccccccc}
    \ddots & \vdots  & \vdots  & \vdots  & \vdots  & \vdots  & \vdots  & \iddots \\
    \dots & M^{m+1}_{+} & V_{\mu_{m+1},\mu_{m+1}+1} & 0 & \tilde{\Delta}^{-}_{\mu_{m+1},\mu_m+1} & 0 & 0 & \dots \\
    \dots & V^{\rm T}_{\mu_{m+1},\mu_{m+1}+1} & M_{-}^{m+1} & (\tilde{\Delta}^{+}_{\mu_m,\mu_{m+1}+1})^{\dag} & 0 & 0 & 0 & \dots \\
    \dots & 0 & \tilde{\Delta}^{+}_{\mu_m,\mu_{m+1}+1} & M_{+}^{m} & V_{\mu_{m},\mu_{m}+1} & 0 & \tilde{\Delta}^{-}_{\mu_{m},\mu_{m-1}+1} & \dots \\
    \dots & (\tilde{\Delta}^{-}_{\mu_{m+1},\mu_m+1})^{\dag} & 0 & V^{\rm T}_{\mu_{m},\mu_{m}+1} & M_{-}^{m} & (\tilde{\Delta}^{+}_{\mu_{m-1},\mu_{m}+1})^{\dag} & 0 & \dots \\
    \dots & 0 & 0 & 0 & \tilde{\Delta}^{+}_{\mu_{m-1},\mu_{m}+1} & M_{+}^{m-1}  & V_{\mu_{m-1},\mu_{m-1}+1} & \dots \\
    \dots & 0 & 0 & (\tilde{\Delta}^{-}_{\mu_{m},\mu_{m-1}+1})^{\dag} & 0 & V^{\rm T}_{\mu_{m-1},\mu_{m-1}+1} & M_{-}^{m-1} & \dots \\
    \iddots & \vdots & \vdots  & \vdots  & \vdots  & \vdots & \vdots & \ddots \\
  \end{array}
\right]\tilde{\phi}_{n,j}
       =E^{n}_{j}\tilde{\phi}_{n,j},\nonumber\\
\end{eqnarray}
where, 

\begin{eqnarray}
&&\tilde{\phi}^{\rm T}_{n,j} =\left[
                          \begin{array}{cccccccc}
                            \dots & \hat{a}^{m+1}_{n,j} & \hat{b}^{m+1}_{n,j} & \hat{a}^{m}_{n,j} & \hat{b}^{m}_{n,j} & \hat{a}^{m-1}_{n,j} & \hat{b}^{m-1}_{n,j} & \dots \\
                          \end{array}
                        \right],
                        \; (\hat{a}^{m}_{n,j})^{\rm T}=e^{-i\pi/4}\left[
                                              \begin{array}{ccccc}
                                                a^{m}_{n,j,1} & \dots&  a^{m}_{n,j,\nu} &  a^{m}_{n,j,\nu+1} & \dots \nonumber\\
                                              \end{array}
                                            \right]
                        \\
&&(\hat{b}^{m}_{n,j})^{\rm T}=e^{i\pi/4}\left[
    \begin{array}{ccccc}
    b^{m}_{n,j,1} & \dots&  b^{m}_{n,j,\nu} &  b^{m}_{n,j,\nu+1}   \end{array}
    \right],\;  [M^{m}_{\pm}]_{\nu,\nu'}=\left[\pm M+\left(m-\frac{1}{2}\right)\hbar\Omega\right]\delta_{\nu,\nu'},\nonumber\\
&& [V_{\mu_m,\mu_{m'}}]_{\nu,\nu'}=\frac{2\hbar v_{\rm F}}{R}\frac{\alpha_{\nu,\mu_m}\alpha_{\nu',\mu_{m'}}}{\alpha^2_{\nu,\mu_m}-\alpha^2_{\nu',\mu_{m'}}},\; [\tilde{\Delta}^{\pm}_{\mu_m,\mu_{m'}}]_{\nu,\nu'}=-i\int_{0}^{R}\phi^{m}_{\nu,n,s,\mu_{m}}(r)\tilde{A}_{\pm}(r)\phi^{m'}_{\nu,n,\bar{s},\mu_{m'}}(r)rdr,
\end{eqnarray} 
$\bar{s}\ne s$, and $s=(\uparrow,\downarrow)$. Similarly, the numerical diagonalization of the radial equations of the projected Floquet Hamiltonian in Eq.~\eqref{EffectiveR} is solved in the $m=\{1,0\}$ subspace of the matrix equation in Eq.~\eqref{HmatL}. 

\subsection{Numerical Solutions of the Mapped BdGEs and Comparison to the Full Floquet Hamiltonian.}
To numerically diagonalize Eq.~\eqref{mappedH} and obtain the bulk spectral properties, similar to the full Floquet Hamiltonian, we use the Bessel decomposition method and we expresses the spinor $\hat{\Psi}(r)$ as
\begin{equation}
\hat{\Psi}(r) = \sum_{\nu=1}^{N}\left[
                 \begin{array}{c}
                   a_{n,j,\nu}\phi^{0}_{\nu,n,\uparrow,j+\frac{\ell}{2}}(r) \\
                   b_{n,j,\nu}\phi^{1}_{\nu,n,\downarrow,j-\frac{\ell}{2}}(r) \\
                 \end{array}
               \right],
\end{equation}
where $\phi^{m}_{\nu,n,\uparrow,j}(r)$ is given in Eq.~\eqref{bessels1}. In this case, we chose $N$ to be the cutoff of $\nu$ which is the number of zeros of the Bessel functions. With this cutoff the size of the matrix to diagonalize at each $j$, is $2N\times2N$, and the matrix equation is written as
\begin{equation}\label{BesselMapped}
\left[
  \begin{array}{cc}
    T_{j+\frac{\ell}{2}} & \Delta_{j+\frac{\ell}{2},j-\frac{\ell}{2}} \\
    \Delta^{\rm T}_{j+\frac{\ell}{2},j-\frac{\ell}{2}} & -T_{j-\frac{\ell}{2}} \\
  \end{array}\right]
\left[
         \begin{array}{c}
           \hat{a}_{n,j} \\
           \hat{b}_{n,j} \\
         \end{array}
       \right]=\mathscr{E}\left[
         \begin{array}{c}
           \hat{a}_{n,j} \\
           \hat{b}_{n,j} \\
         \end{array}
       \right],
\end{equation}
where $\mathscr{E}=E^{n}_{j}$, 
\begin{eqnarray}
&&(\hat{a}^{m}_{n,j})^{\rm T}=\left[
\begin{array}{ccccc}
a^{m}_{n,j,1} & \dots&  a^{m}_{n,j,\nu} &  a^{m}_{n,j,\nu+1} & \dots \\
\end{array}
\right],\;  
(\hat{b}^{m}_{n,j})^{\rm T}=\left[
\begin{array}{ccccc}
b^{m}_{n,j,1} & \dots&  b^{m}_{n,j,\nu} &  b^{m}_{n,j,\nu+1}   &\dots \end{array}
\right],  \\
&&[T_{j}]_{\nu,\nu'}= \frac{\hbar^2}{2m^{*}}\left[\left(\frac{\alpha_{\nu,j}}{R}\right)^2-k^2_{\rm F}\right]\delta_{\nu,\nu'},\;{\rm and}\; [ \Delta_{j+\frac{\ell}{2},j-\frac{\ell}{2}}]_{\nu,\nu'}=\int_{0}^{R}\phi^{0}_{\nu,n,\uparrow,j+\frac{\ell}{2}}(r)\Delta(r)\phi^{1}_{\nu',n,\downarrow,j-\frac{\ell}{2}}(r)rdr.
\end{eqnarray}
\subsubsection{Comparison to the Full Floquet Hamiltonian}
In what follows, we compare the bulk quasienergy spectrum of the VLB irradiated system obtained from diagonalizing the Floquet Hamiltonian with $N_{\rm F}$ Floquet copies [Eq.~\eqref{HmatL}] with the spectrum obtained from the mapped BdGEs in Eq.~\eqref{BesselMapped}.  

\begin{figure}[ht!]
  \centering
  \includegraphics[width=1\textwidth]{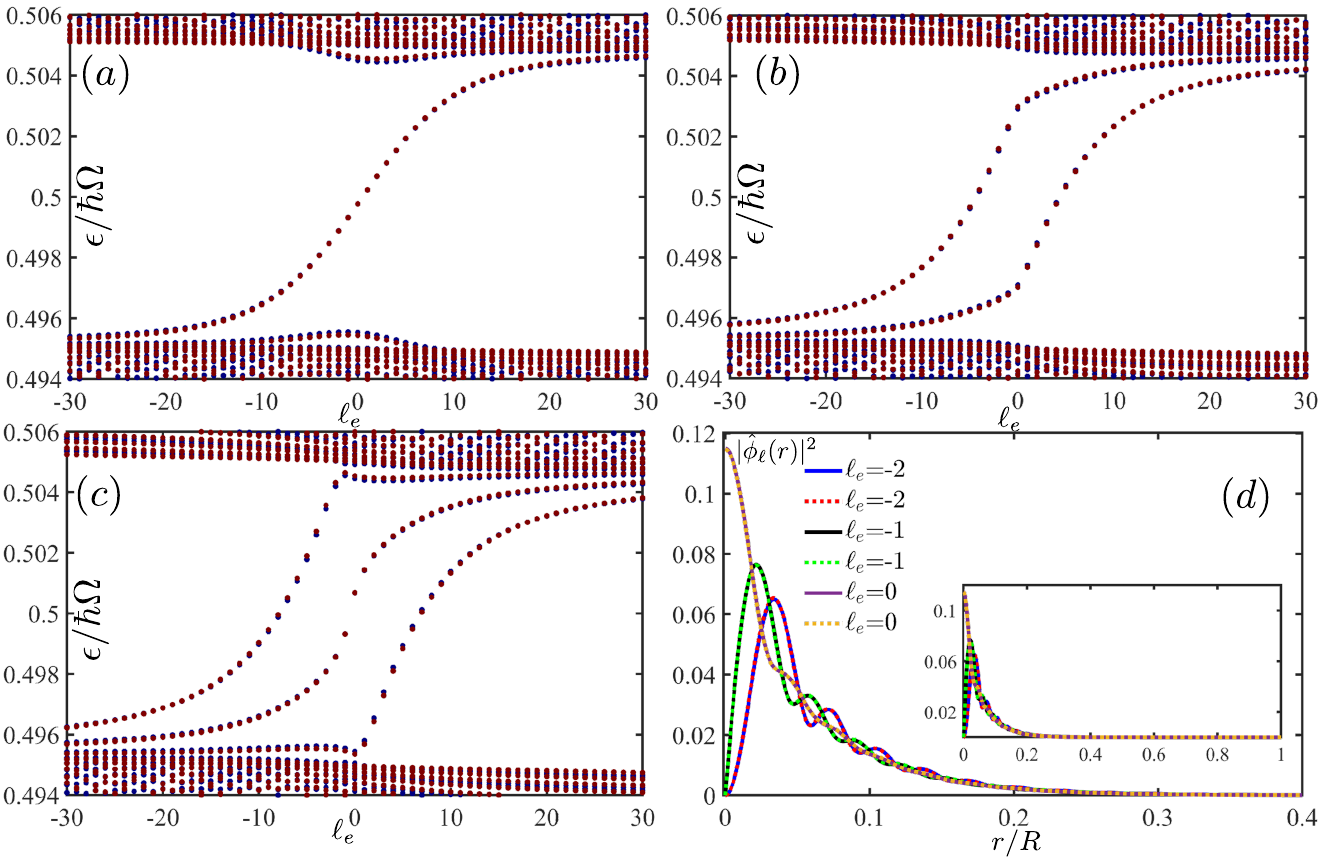}
  \caption{Comparison of the bulk quasienergy spectra obtained from the numerical diagonalization of the Floquet Hamiltonian for the VLB driven system, Eq.~\eqref{HmatL}, and the BdGEs describing the same system, Eq.~\eqref{BesselMapped}. (a)-(c) Correspond to a VLB carrying orbital angular momentum $\ell=-1,-2$ and $-3$, respectively. The quasienergies obtained from Eq.~\eqref{HmatL} are represented with the blue dots, and were obtained by considering $N_{\rm F}=5$ and $N=250$. The eigenvalues obtained from Eq.~\eqref{BesselMapped} are represented with the red dots. (d) The radial extension of the vortex states with $\ell_e=-2,-1$, and $0$ for a VLB with $\ell=-1$, obtained from the Floquet Hamiltonian (solid lines), Eq.~\eqref{HmatL}, and the effective BdGEs (dotted lines), Eq.~\eqref{BesselMapped}. The inset of (d) shows the extension of the states over the full sample.  We also note that the driven system's parameters are $\hbar\Omega=2.06M$, $k_{\rm F}\xi=10$, $\Delta_0=0.01M$, $\Delta(r)=\Delta_0\tanh(r/\xi)$, $R=10\xi$, $v_{\rm F}/\Omega=1$\AA, and $2M$ is the semiconducting gap of the Dirac material [same parameters as in Fig. (2) in the main text].}
  \label{figs}
\end{figure}
The inspection of panels (a)-(c) in Fig.~\ref{figs} reveals an excellent agreement between the diagonalization of a Floquet Hamiltonian with $m=\{2,1,0,-1,-2\}$ Floquet sidebands, and the effective BdGEs in the regime defined by frequency $\hbar\Omega \gtrsim 2M$ and with a light matter coupling $g=ev_{\rm F}A_{0}/(\hbar\Omega)\ll 1$. The weak light matter coupling in the system ensures that at the edge of the Floquet Brillouin zone the dominant contributions are due to the $m=1$ and $0$ Floquet bands and that additional anti-crossings do not impact the quasienergy. We also notice that for the light matter parameters and sample size considered, Fig.~\ref{figs}, $N>200$ is sufficient for obtaining consistent numerical results for the spectrum shown in Fig.~\ref{figs} (a)-(c), and states shown in Fig.~\ref{figs} (d). Additionally, the highest relative difference between the vortex states quasienergies obtained by Eq.~\eqref{HmatL} and~\eqref{BesselMapped} for $\ell=-1$ is of the order $\sim 10^{-5}$, while for $\ell=-2\;{\rm and}\; -3$ it is of the order $\sim 10^{-4}$. 

The real space extension of the vortex states is shown in Fig.~\ref{figs} (d) and its inset, for $\ell=-1$. Since these states have azimuthal symmetry, we show their radial dependence. From this figure, one can notice that these states are highly localized at radii $R_{0}\approx0$, $R_{-1}\approx R/50$, and $R_{-2}\approx 3R/100$ for $\ell_e=0,-1\; {\rm and}\; -2$. We also note that the spatial extension of the states with $\ell_e=(-2,-1,0)$ is identical to that of the states with $\ell_e=(1,2,3)$, and this is reflective of the symmetry $\mathcal{S}_1: \mathscr{E}^{\rm V}_{j+1+\frac{\ell}{2}}=-\mathscr{E}^{\rm V}_{-j+1+\frac{\ell}{2}}$. Finally we notice an excellent agreement between the states obtained from the Floquet Hamiltonian and those obtained in through the BdEGs, indicating that accurate description of the VLB driven system provided by the effective BdEGs.      
\end{widetext}     

\end{document}